\documentclass[]{pasj01}

\Received{2023/09/29}
\Accepted{2023/12/29}
 
\usepackage{hyphenat} 
\usepackage[switch,mathlines]{lineno}
\usepackage{url}
\usepackage{comment}
\usepackage{upgreek}
\usepackage{diagbox}
 
\begin{document} 

\title{ 
GALAXY CRUISE: Spiral and ring classifications for bright galaxies at z=0.01--0.3}

\author{Rhythm \textsc{Shimakawa}\altaffilmark{1,2}}
\altaffiltext{1}{Waseda Institute for Advanced Study (WIAS), Waseda University, Nishi Waseda, Shinjuku, Tokyo 169-0051, Japan}
\altaffiltext{2}{Center for Data Science, Waseda University, 1-6-1, Nishi-Waseda, Shinjuku, Tokyo 169-0051, Japan}
\email{rhythm.shimakawa@aoni.waseda.jp (RS)}

\author{Masayuki \textsc{Tanaka}\altaffilmark{3}}
\altaffiltext{3}{National Astronomical Observatory of Japan (NAOJ), National Institutes of Natural Sciences, Osawa, Mitaka, Tokyo 181-8588, Japan}

\author{Kei \textsc{Ito}\altaffilmark{4}}
\altaffiltext{4}{Department of Astronomy, Graduate School of Science, The University of Tokyo, 7-3-1 Hongo, Bunkyo, Tokyo 113-0033, Japan}

\author{Makoto \textsc{Ando}\altaffilmark{4}}




\maketitle

\begin{abstract}
This paper presents a morphology classification catalog of spiral and ring features of 59,854 magnitude-limited galaxies ($r<17.8$ mag, and additional 628,005 subsamples down to $r=20$ mag) at $z=$ 0.01--0.3 based on the Third Public Data Release of the Hyper Suprime-Cam Subaru Strategic Program.
We employ two deep learning classifiers to determine the spiral and ring structures separately based on GALAXY CRUISE Data Release 1, which is dedicated to Hyper Suprime-Cam data.
The number of spiral and ring galaxies contain 31,864 and 8,808 sources, respectively, which constitute 53\% and 15\% of the sample. 
A notable result of this study is the construction of a large sample of ring galaxies utilizing high-quality imaging data delivered by the Subaru Hyper Suprime-Cam. 
However, the accurate identification of ring galaxies remains difficult at a limited seeing resolution.
Additionally, we confirm that most spiral galaxies are located on the star-forming main sequence, whereas ring galaxies preferentially reside in the green valley at stellar mass of $10^{10.5}$--$10^{11}$ solar mass.
Furthermore, decreasing fractions of spiral and ring galaxies are observed toward the centers of the galaxy clusters.
The obtained morphology catalog is publicly available on the GALAXY CRUISE website.
\end{abstract}


\section{Introduction}

The vast expanse of the universe houses myriad galaxies featuring diverse morphologies \citep{Hubble1926,Sandage1961,Willett2013,Tanaka2023}. 
Galaxies from near and far have been investigated comprehensively to systematically understand the formation mechanisms and diversity of their morphologies.

Spiral arms are known as a representative feature that characterizes the morphological types of galaxies \citep{Lin1964,Goldreich1965,Sellwood1984,Bertin1989,DOnghia2013}.
Spiral galaxies, which is characterized by their prominent spiral arms, have intrigued astronomers since their discovery in the late 18th century using a 72-inch telescope by William Parsons, the third Earl of Rosse \citep{Seigar2017}.
Currently, the formation mechanism of spiral galaxies have been duly considered, particularly from the theoretical perspective, but remain unsettled (see reviews by \cite{Dobbs2014,Sellwood2022} and references therein). 
Investigations into spiral galaxies have yielded invaluable insights into the nature of galactic dynamics, star formation, and the evolution of cosmic structures.
Observational studies have demonstrated correlations between the pitch angle and mass concentration, colors and black hole mass \citep{Seigar2008,Berrier2013,Kendall2015,Willett2015,Davis2017,Hart2017,Yu2019,Yu2022}, and the morphology--density relation (e.g., \cite{Dressler1980,Dressler1997,Larson1980,Goto2003,Postman2005,Smith2005,Bamford2009,Cappellari2011,Mei2023}).
Thus, spiral arms are important for unraveling the formation history of galaxies.

Similarly intriguing are ring galaxies, a subclass of spiral galaxies with concentric rings encircling their nuclei \citep{Hoag1950,Buta1996}. 
These celestial gems offer insights into the complex relationship among gravitational forces, stellar interactions, and the dynamics of galactic collisions. 
Subsequent observations and studies have revealed various ring galaxies, which were categorized into the crisp P-type and smooth O-type \citep{Few1986}, each of which exhibited unique characteristics and evolutionary paths \citep{Fosbury1977,Schweizer1983,Buta1990,Elmegreen1992,Appleton1996,Grouchy2010,Herrera-Endoqui2015,Buta2017,Fernandez2021}. 
Theories propose that they may arise from encounters between galaxies, tidal interactions with neighboring galaxies, or gravitational effects induced by a central galactic bar \citep{Lynds1976,Theys1976,Theys1977,Schwarz1981,Struck-Marcell1990,Romano2008,Elagali2018,Murugeshan2023,Tous2023}. 
However, ring galaxies require more research than spiral galaxies owing to their scarcity and identification difficulties \citep{Fernandez2021}.
Notably, discerning ring galaxies from pseudo-ring galaxies using seeing-limited images from ground-based telescopes is challenging (e.g., \cite{Shimakawa2022}).

Statistical surveys of these substructures are important for unraveling the formation histories of the galaxy morphology and its diversity in the universe. 
In particular, a large sample of ring galaxies was not constructed, until recently \citep{Walmsley2022}, owing to various issues in determining the ring features. 
Advancements in technological capabilities have revolutionized imaging quality and area coverage, as represented by the Subaru Hyper Suprime-Cam \citep{Miyazaki2018}, Vera C. Rubin Observatory \citep{Ivezic2019}, and Euclid \citep{Laureijs2011,Amendola2013}, which provide unprecedented opportunities for the investigation of spiral and ring galaxies. 
Specifically, imaging data from the Hyper Suprime-Cam Subaru Strategic Program (HSC-SSP; \cite{Aihara2018}) feature approximately twice the spatial resolution of the Sloan Digital Sky Survey (SDSS; \cite{York2000,Strauss2002,Gunn2006}) and the Panoramic Survey Telescope and Rapid Response System (Pan-STARRS; \cite{Chambers2016,Flewelling2020}) and are approximately by two mag deeper than the Dark Energy Survey (DES; \cite{Flaugher2015,Abbott2021}). 
However, their survey areas are significantly larger than that of the HSC-SSP.
Indeed, various previous studies have demonstrated that the HSC-SSP data are quite useful for large-scale morphological classifications \citep{Tadaki2020,Martin2020,Shimakawa2021,Shimakawa2022,Ghosh2023,DominguezSanchez2023,Omori2023}.

Based on the background above, this study is conducted to construct a morphology catalog of spiral and ring galaxies based on HSC-SSP Public Data Release 3 (HSC-SSP PDR3; \cite{Aihara2022}).
Deep and high-resolution images from the HSC-SSP enable comprehensive investigations into these galactic structures, thus facilitating the acquisition of practical datasets for developing theoretical frameworks \citep{Bottrell2023,Eisert2023}.
However, definitely identifying of ring galaxies remains challenging, even when using HSC-SSP data, as reported by \citet{Shimakawa2022}.
Moreover, establishing training data for ring galaxies is more expensive compared with that for spiral galaxies owing to the rarity of ring galaxies.
Therefore, we utilize the visual classifications by GALAXY CRUISE \citep{Tanaka2023} as labeled training data for deep learning classifiers. 
GALAXY CRUISE is a citizen science project dedicated to HSC-SSP imaging data\footnote{\url{https://galaxycruise.mtk.nao.ac.jp/en/}}, which allows us to maximize its great imaging quality and hence identify suitable ring galaxy candidates.

The remainder of this paper is organized as follows: First, we provide an overview of the sample selection and image preprocessing based on SDSS DR17 \citep{Blanton2017,Abdurrouf2022} and HSC-SSP PDR3 \citep{Aihara2022} in section~\ref{s2}.
Section~\ref{s3} presents the approach adopted to establish the deep learning classifiers to select spiral and ring galaxies at $z=$ 0.01--0.3 using labeled training data from GALAXY CRUISE Data Release 1 (GC~DR1; \cite{Tanaka2023}). 
Subsequently, we analyze the stellar masses and star formation rates (SFRs) of the obtained spiral and ring galaxies and their environmental dependence using value-added public catalogs (section~\ref{s4}).
Finally, we summarize our results and discussion in section~\ref{s5}.
The AB magnitude system \citep{Oke1983} and the \citet{Chabrier2003} stellar initial mass function (IMF) are adopted in this study. 
Additionally, a flat WMAP7 cosmology ($H_0=70$ km~s$^{-1}$Mpc$^{-1}$, $\Omega_m=0.27$; \cite{Komatsu2011}) is assumed.

\section{Data preparation}\label{s2}

The primary goal of this study is to apply a labeled training dataset from the GC~DR1 to the publicly available HSC-SSP images of local bright galaxies.
We adopted a sample comprising 687,859 magnitude-limited ($r<20$) galaxies at $z=$ 0.01--0.3, which are eventually divided into 59,854 main sample ($r<17.8$) and 628,005 subsample ($r=$ 17.8--20 mag) as shown in table~\ref{tab1}.
The final dataset was constructed based on multiple selection criteria and cross-matching using photometric information and spectroscopic redshifts from SDSS DR17 \citep{Blanton2017,Abdurrouf2022}, and imaging data from the HSC-SSP PDR3 Wide Layer (\cite{Aihara2022}, hereinafter referred to as HSC-SSP PDR3).
This section explains the establishment of the dataset and discusses the potential sampling bias.

\subsection{Target selection}\label{s21}

\begin{figure*}
\begin{center}
\includegraphics[width=14cm]{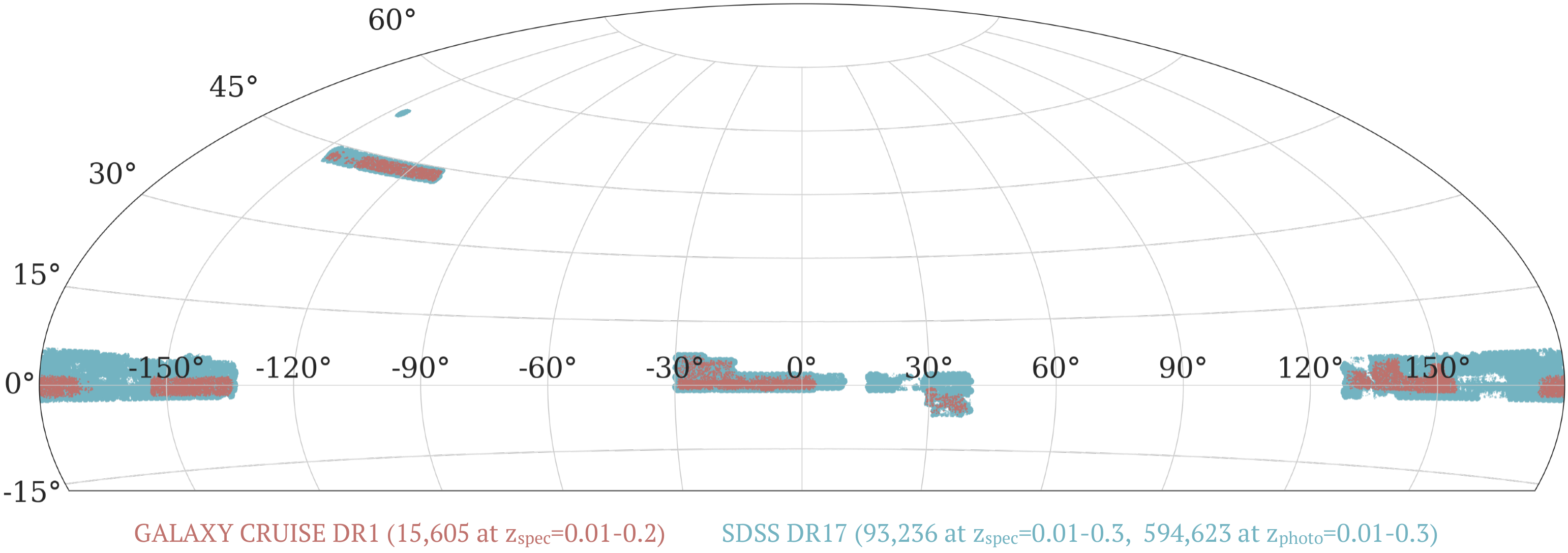}
\end{center}
\caption{
Sky distributions of our samples.
Red and blue symbols show labeled training data with GC~DR1 at $z=$ 0.01--0.2 and our targets at $z=$ 0.01--0.3 from HSC--SDSS database, respectively.
}
\label{fig1}
\end{figure*}

We first prepared magnitude-limited objects with spectroscopic or photometric redshifts from SDSS DR17 \citep{Blanton2017,Abdurrouf2022} based on the selection criteria below:
\begin{itemize}
   \item Spec-$z$ sample ($N\sim920,000$)
   \begin{itemize}
     \item $z=$ 0.01--0.30
     \item $r_\mathrm{petro}<20$ mag
     \item petrosian $r$ radius (R90) $<30$ arcsec
     \item clean photometry flag $=1$
     \item redshift error flag $=0$
   \end{itemize}
   \item Photo-$z$ sample ($N\sim8,200,000$)
   \begin{itemize}
     \item $z=$ 0.01--0.30
     \item $r_\mathrm{petro}=$ 17.8--20.0 mag
     \item petrosian $r$ radius (R90) $<30$ arcsec
     \item clean photometry flag $=1$
     \item photometric error class $=1$ (best photo-$z$ class)
   \end{itemize}
\end{itemize}
We do not employ bright photo-$z$ sources with $r_\mathrm{petro}<17.8$ mag to avoid the foreground/background contamination to the SDSS Main Galaxy Sample \citep{Strauss2002}, which spectroscopically cover the $r$-band limited sample ($r_\mathrm{petro}<17.8$ mag) at 90--95\% completeness rate.
We set the apparent size limit using the Petrosian radius (R90 $<30$ arcsec) in the $r$-band, where R90 is defined as the radius containing 90\% of the Petrosian aperture flux \citep{Abdurrouf2022} to remove extremely large objects.
We disregarded these large objects ($<0.1\%$ of the total) because they were excessively resolved and large for the current HSC-SSP pipeline, which implies that they tend to be affected from sky subtraction and deblending problems.
Section~\ref{s22} describes the fraction of galaxies missed owing to the size limit and cross-matching process, as mentioned below.

Subsequently, we cross-matched these samples with the HSC-SSP PDR3 sources detected in the $i$-band within a radius of 1 arcsec.
It should be noted that the the HSC-SSP PDR3 cover only a narrow slice ($\sim7\%$) of the imaging sky coverage of 14,555 deg$^2$ in the SDSS DR17 \citep{Abdurrouf2022}.
These HSC sources were selected using multiple quality flags for quality assessments to select clean objects, as described below:
\begin{description}
   \item[--] {\tt isprimary is True}
   \item[--] {\tt i\_sdssshape\_flag\_badcentroid is False}
   \item[--] {\tt i\_psfflux\_flag is False}
   \item[--] {\tt i\_pixelflags\_edge is False}
   \item[--] {\tt i\_pixelflags\_bad is False}
   \item[--] {\tt i\_pixelflags\_crcenter is False}
   \item[--] {\tt i\_inputcount\_flag\_noinputs is False}
   \item[--] {\tt i\_mask\_brightstar\_halo is False}
   \item[--] {\tt i\_mask\_brightstar\_ghost is False}
   \item[--] {\tt i\_mask\_brightstar\_blooming is False}
\end{description}
The details of these catalog flags are described by \citet{Coupon2018,Bosch2018,Aihara2022}.
We applied the seeing cut of full width at half-maximum (FWHM) $<0.7$ arcsec in the $i$-band to exclude low-quality images, which further reduced the sample size by 10\%.
Consequently, 93,236 spec-$z$ sources and 594,623 photo-$z$ sources in HSC-SSP PDR3 were successfully matched to the SDSS samples without duplication (table~\ref{tab1}).
Their sky distributions are shown in figure~\ref{fig1}.

\begin{table}
\tbl{The target list of this work.
Here, galaxies with model predictions of {\tt prob\_sp} and {\tt prob\_rg} $>0.5$ (table~\ref{tab3}) are defined as spiral and ring galaxies, respectively.}{%
\begin{tabular}{lrrr}
    \hline
    Name              & N$_\mathrm{total}$ & N$_\mathrm{spiral}$ & N$_\mathrm{ring}$\\
    \hline
    GALAXY CRUISE DR1 &  15,605  & $^\ast$6,038 &  $^\ast$930\\
    \hline
    HSC-SSP PDR3            & 687,859  & 385,449   &  33,993\\
    Main sample ($r<17.8$)  &  59,854  &  31,864   &   8,808\\
    Subsample ($r=17.8-20$) & 628,005  & 353,585   &  25,185\\
    \hline
    \end{tabular}}
\label{tab1}
\begin{tabnote}
\footnotemark[$\ast$] Number of sources based on relatively conservative selections for training datasets (see text).\\ 
\end{tabnote}
\end{table}

Additionally, we employed the morphological classification labels from the GC~DR1 catalog (see \cite{Tanaka2023} for the complete details).
The original target selection in GALAXY CRUISE is based on the HSC-SSP PDR2 samples \citep{Aihara2019} with spectroscopic confirmations at $z=$ 0.01--0.2 by either or both of SDSS DR15 \citep{Aguado2019} and the GAMA survey (DR2; \cite{Driver2009,Liske2015}). 
Thus, all GC~DR1 sources (N=20,686) were included by the survey area of HSC-SSP PDR3 \citep{Aihara2022}, as shown in figure~\ref{fig1}.
However, the aforementioned selection criteria reduced the sample size to 15,606 objects.
Table~\ref{tab1} summarizes the targets used in this study.

\begin{figure}
\begin{center}
\includegraphics[width=7.5cm]{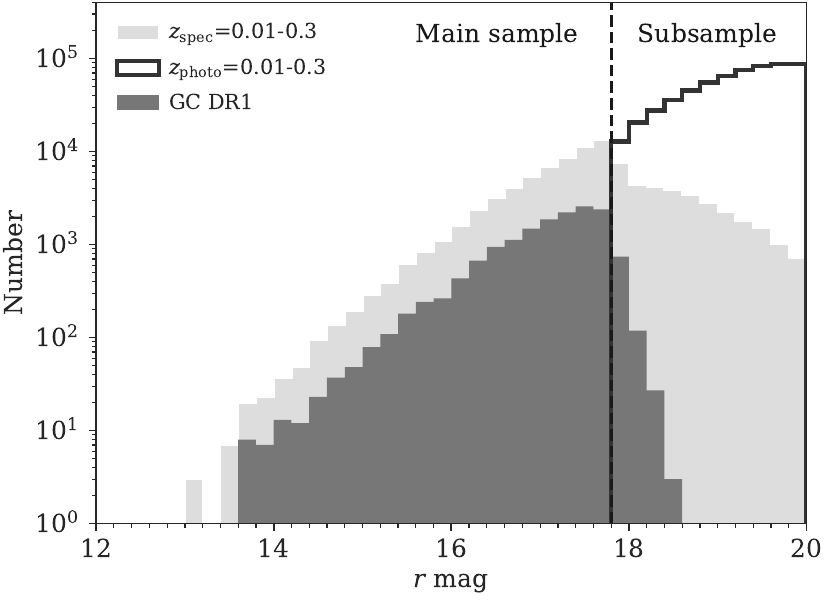}
\end{center}
\caption{
Number distribution as a function of $r$-band magnitude.
Light and dark gray histograms indicate those of entire spec-$z$ sample and training sample with morphology information from the GC~DR1, respectively.
We employed photo-$z$ sources (opened histogram) to complement spec-$z$ sample at magnitude range of $r=$ 17.8--20.0 mag.
The whole sample is eventually divided into the main sample consisting of purely magnitude-limited ($r<17.8$) galaxies with spectroscopic redshifts, and the subsample with $r=$ 17.8--20 mag (table~\ref{tab1}).
}
\label{fig2}
\end{figure}

Figure~\ref{fig2} shows the $r$-band magnitude distributions of our scientific targets and GC~DR1 sources.
Because the GC~DR1 sources are primarily based on the SDSS Main Galaxy Sample \citep{Strauss2002}, which is a purely magnitude-limited sample ($r<17.8$), they are primarily distributed at $r\lesssim18$ mag.
Therefore, we decided to use $r$-band magnitude-limited galaxies with $r<17.8$ mag as our main targets (table~\ref{tab1}).
The remaining with $r=$ 17.8--20 mag are defined as the subsample, which are also applied to our morphological classifications but not used in discussion (section~\ref{s4}).
Because the subsample largely cover targets in the GALAXY CRUISE Season~2 currently in operation, their morphological information will help us to instantly compare citizen's classifications with machine-learning results from this work, which will further promote our outreach activities in the near future.

\subsection{Sampling bias}\label{s22}

\begin{figure}
\begin{center}
\includegraphics[width=7cm]{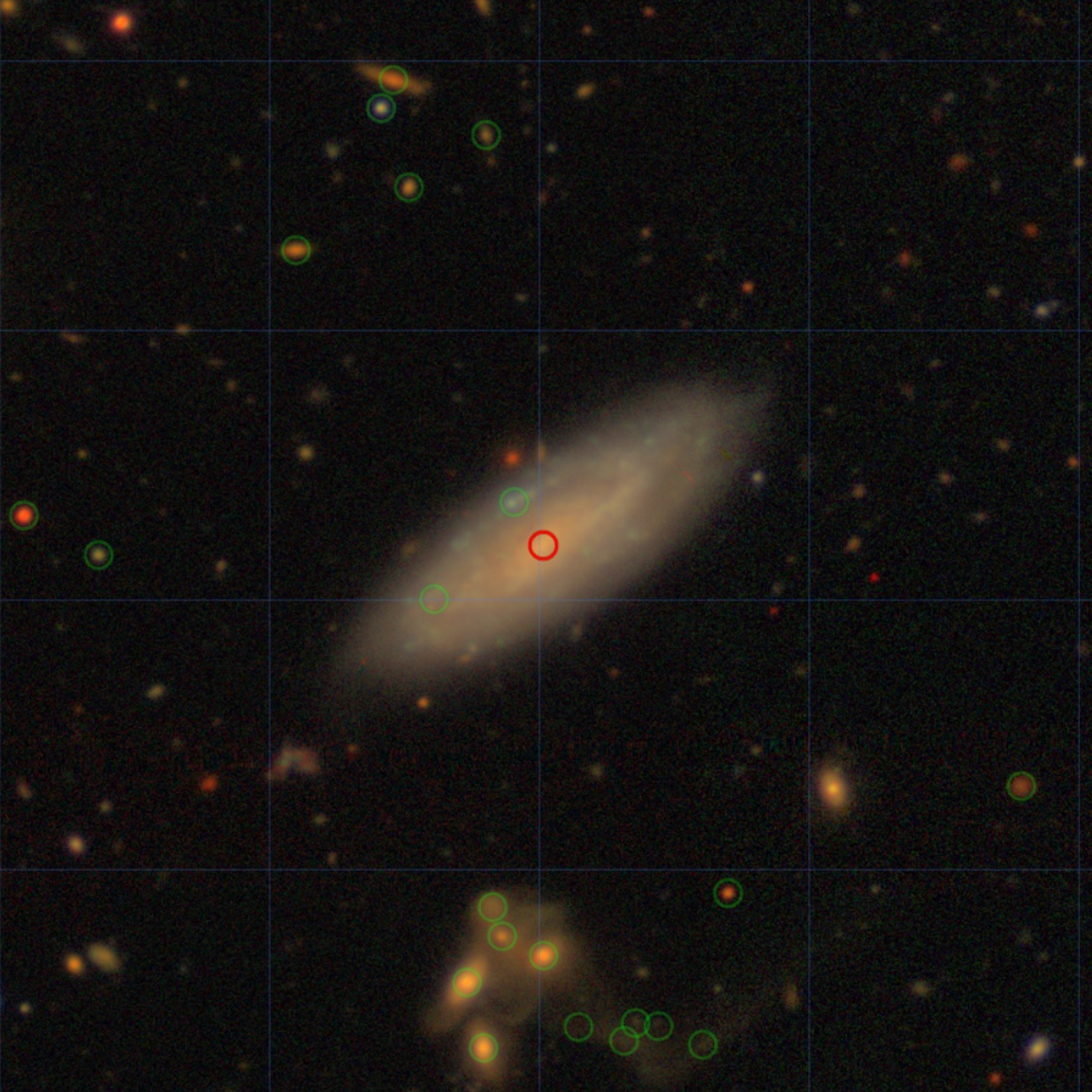}
\end{center}
\caption{
Example image of the mismatched SDSS sources on the {\tt hscMap}\footnote{\url{https://hsc-release.mt k.nao.ac.jp/hscMap-pdr3/app}}.
Central source denoted by the red circle comprises a bright galaxy from SDSS DR17 but failed to cross-match with HSC-SSP PDR3 sources due to the failure of centroid detection ({\tt i\_sdssshape\_flag\_badcentroid is True}).
Green circles are $i$-band magnitude limited sources ($i<22$) satisfying the selection criteria for HSC-SSP PDR3, which show misidentifications in extended disk and tidal features.
}
\label{fig3}
\end{figure}

Although the $r$-band magnitude-limited sources in the SDSS DR17 database were cross-matched with those in the HSC-SSP PDR3 almost completely, some sources did not belong to the sample because of the detection failures of centroids, the size criteria (R90 $<30$ arcsec; section~\ref{s21}), and deblending.
Deblending occured in the source detection process, particularly on the HSC side, because some local bright galaxies at $z<0.07$ were ``excessively resolved" and thus misinterpreted as multiple sources, not an identical object (see an example in figure~\ref{fig3}). 

\begin{figure}
\begin{center}
\includegraphics[width=7.5cm]{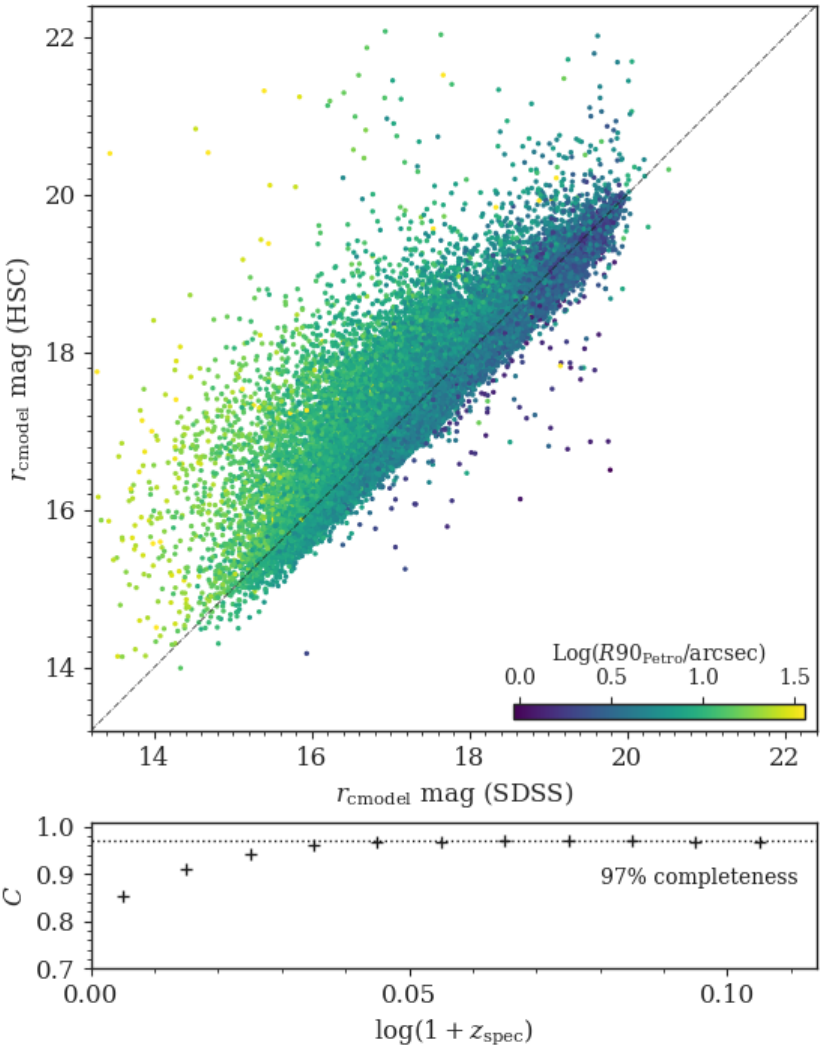}
\end{center}
\caption{
Top panel: Comparison of Cmodel magnitudes in the $r$-band from SDSS DR17 and HSC-SSP PDR3 for control spec-$z$ sources at $z=$ 0.01--0.03  ($N\sim60,000$).
They are located in a high effective area ($>90$\%), where problematic regions due to such as ghosts and halos of bright stars are minimized to less than 10\%.
Symbol colors indicate Petrosian $r$-band radius (R90), as denoted in the inset color-bar. 
Dash-dotted line depicts the 1:1 relation.
Bottom panel: Success rate of cross-matching ($C$) between SDSS DR17 and HSC-SSP PDR3 as a function of redshift in control field.
Approximately 97\% (the dotted line) of SDSS DR17 sources are successfully matched to our samples from HSC-SSP PDR3, whereas many sources appeared at lower redshifts (i.e., $z<0.07$) due to deblending (figure~\ref{fig3}).
}
\label{fig4}
\end{figure}

To evaluate the fraction of SDSS samples excluded during the selection process, we calculated the cross-match rate ($C$) under a redshift in the HSC-SSP field (figure~\ref{fig4}). 
The matching rate $C$ is defined as the number ratio of SDSS DR17 sources with counterparts in HSC-SSP PDR3 that satisfy the selection criteria (section~\ref{s21}) to the entire sample.
Here, we estimated the $C$ values only for the clean fields with effective areas exceeding 90\%, in which contaminants, such as bright stars and satellite trails should be less than 10\% (see \cite[section~3]{Shimakawa2021} for more details regarding the calculation procedure).
We confirmed that approximately 97\% of the SDSS-selected sources were successfully matched to the HSC-SSP sources down to $z\sim0.07$. 
However, the $C$ value decreased to 85\% at $z\lesssim0.02$ (figure~\ref{fig5}). 
Owing to this deblending issue, we observed significant flux losses for more extended sources via HSC photometry (figure~\ref{fig4}).
Therefore, to avoid the flux loss issue, we decided not to use HSC photometry to measure the physical properties of the targets throughout this study.

\subsection{Data preprocessing}

\begin{figure*}
\begin{center}
\includegraphics[width=15cm]{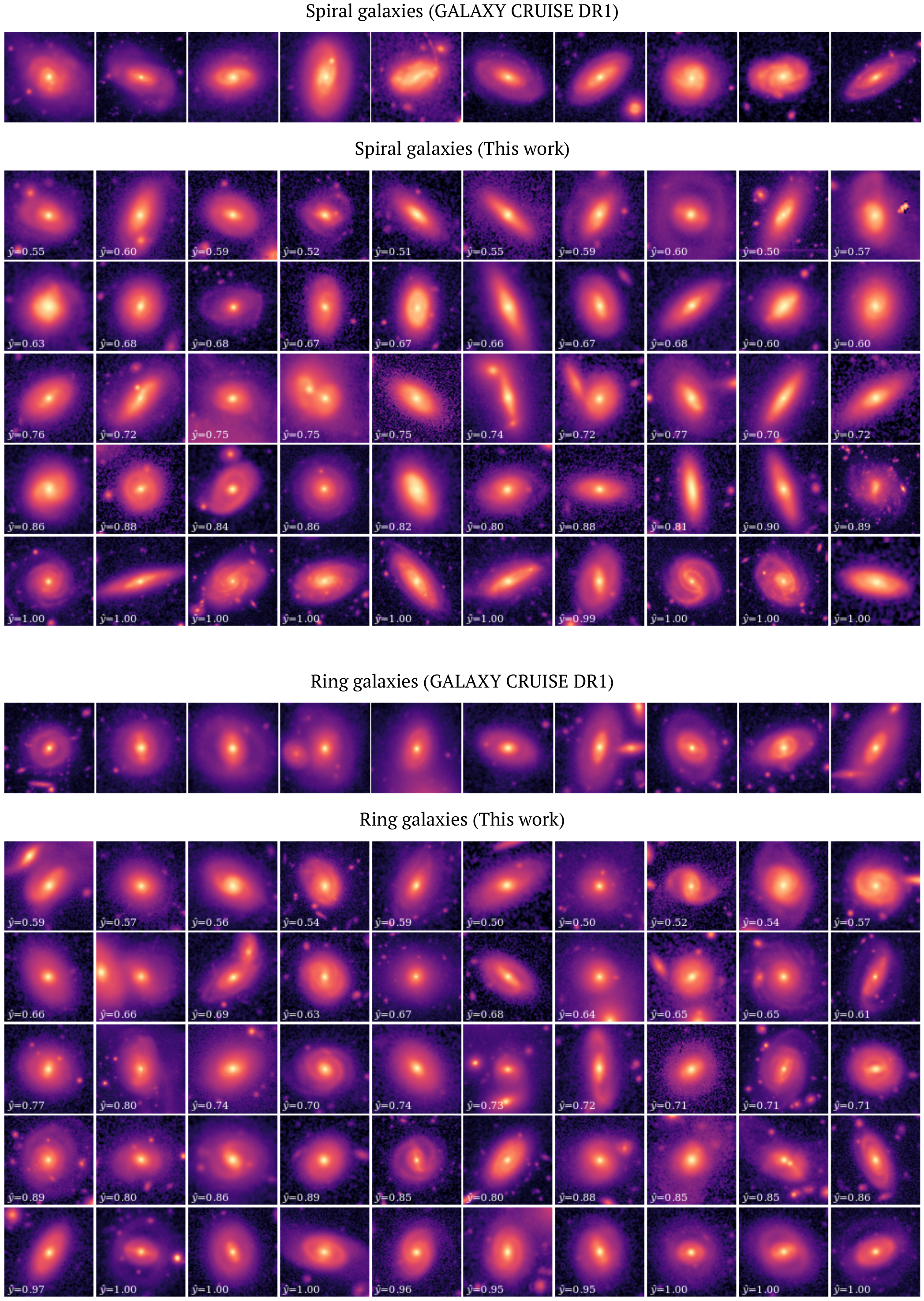}
\end{center}
\caption{
From top to bottom, randomly selected cutout images of spiral galaxies for model training (1st row), those detected from the trained model with $\hat{y}>0.5$ (2nd--6th rows), ring galaxies for the model training (7th row), and those detected from the trained model (8th--12th rows), respectively.
Number corresponding to each image except for training images represents probability ($\hat{y}$) from deep learning classifiers.
}
\label{fig5}
\end{figure*}

We constructed grayscale images in the {\tt png} format in the same manner as in our previous study \citep{Shimakawa2022}, except that the $i$-band had a matched seeing FWHM of 0.7 arcsec.
The following section outlines the image preparation and preprocessing. 
Individual images measuring $(4\times\mathrm{R90})^2$ arcsec$^2$ were cropped from the patches.
The seeing resolution was matched to an FWHM of 0.7 arcsec via Gaussian smoothing, where the images with seeing sizes larger than this threshold were removed, as mentioned in section~\ref{s21}.
The galactic absorption was corrected following \citet{Schlegel1998}.
Image fluxes were normalized using the inverse hyperbolic sine function, arcsinh \citep{Lupton1999,Lupton2004}, which is typically used in image preprocessing for citizen science projects and machine learning classifications in the local universe (e.g., \cite{Lintott2008,Lintott2011,Willett2013}), including the GALAXY CRUISE project \citep{Tanaka2023}.
The flux normalization code was developed based on data access tools provided by the HSC-SSP pipeline team\footnote{\url{https://hsc.mtk.nao.ac.jp/ssp/data-release/}}.
Examples of the constructed images with two different classification types based on the GC~DR1 (spiral and ring galaxies) are shown in figure~\ref{fig5}.

\section{Morphological classification}
\label{s3}

\subsection{Deep learning model}
\label{s31}

\subsubsection{Model architecture}
\label{s311}

As our classification scheme is similar to that of \citet{Shimakawa2021,Shimakawa2022}, we provide only an overview of it below.
We applied the transfer learning \citep{Pratt1992,Oquab2014} to a pre-trained deep convolutional network (CNN; \cite{Lecun1998,Lecun2015}) with the labeled dataset from GALAXY CRUISE (table~\ref{tab1}) for spiral and ring classifications.
All science images were rescaled to $128\times128$ pixels, which corresponded approximately to the typical size of the cutout images of the spec-$z$ sources.
We employed a two-class sigmoid classifier attached to the {\tt Xception} architecture \citep{Chollet2016} using the {\tt Keras} library (version 2.8.0; \cite{Chollet2015}) on {\tt TensorFlow} (version 2.8.2; \cite{Abadi2016}), which was pre-trained with the ImageNet dataset \citep{Russakovsky2014}.
The sigmoid classifier comprised a dense layer ($d=1024$), batch normalization \citep{Ioffe2015}, and ReLU activation \citep{Glorot2011}, which yielded a probability of ($\hat{y}=[0:1]$) for spiral (or ring) galaxies.
The model training and classifications described in the following sections were performed using a single GPU (NVIDIA RTX 6000 Ada).

\subsubsection{Spiral classification}
\label{s312}

We trained the model using the labeled data from the GC~DR1 source catalog \citep{Tanaka2023}.
First, we prepared 10,000 (5,000 spiral and 5,000 non-spiral galaxies) labeled training data for spiral classification.
Each morphological class satisfied the following criteria:
\begin{description}
\item[--] Spiral: {\tt P(spiral)} $>0.79$ $\to$ N$_\mathrm{positive}$=6,038
\item[--] Non-spiral: {\tt P(spiral)} $<0.5$ $\to$ N$_\mathrm{negative}$=6,912,
\end{description}
where {\tt galaxy\_type} is the vote fraction corresponding to the presence of spiral arms in the GALAXY CRUISE; and N$_\mathrm{positive}$ and N$_\mathrm{negative}$ indicate the number of sources that satisfy the selection requirements for the spiral and non-spiral types, respectively.
We adopted a relatively conservative threshold of {\tt galaxy\_type} $>0.79$ to select spiral galaxies according to a suggestion by \citet{Tanaka2023}, and then randomly selected 5,000 from 6,038 sources satisfying this criterion.
We should note that our conclusion does not change even if we adopt {\tt galaxy\_type} $>0.5$ to select spiral galaxies as it has a minor impact on the training dataset due to the bimodal distribution of the vote fraction \citep{Tanaka2023}.
Meanwhile, we selected 5,000 objects with {\tt galaxy\_type} $<0.5$ (N$_\mathrm{negative}$=6,912) as non-spiral galaxies.
In each class, we selected 1,000 spiral (and no-spiral) galaxies as labeled test data from the remaining sources to independently evaluate the model performance (table~\ref{tab2}).

We trained the model until the binary cross-entropy loss of the validation data no longer reached a minimum across more than 10 epochs (known as ``early stopping") in K-folds cross-validation with $K=5$. 
Thus, 8,000 and 2,000 images were used for model training and validation, respectively.
Similar to previous studies \citep{Shimakawa2021,Shimakawa2022}, these images were transformed and augmented using random horizontal and vertical flips and rotations in each epoch.
Parameter tuning was conducted using an adaptive learning rate optimization algorithm ({\tt Adam}; \cite{Kingma2014}) through stochastic mini-batch sampling with a batch size of 32.
Consequently, we obtained the best model with the least validation loss (0.0002) and the best validation accuracy (0.98) at the 14th epoch. 

Evaluation metrics of the best model in the independent test data (N=2,000) are summarized in table~\ref{tab2}, in which the three metrics mentioned above are defined as follows:
\begin{eqnarray} 
  \mathrm{Accuracy} &=& \frac{\mathrm{TP+TN}}{\mathrm{TP+TN+FP+FN}} \\
  \mathrm{Precision} &=& \frac{\mathrm{TP}}{\mathrm{TP+FP}} \\
  \mathrm{Recall} &=& \frac{\mathrm{TP}}{\mathrm{TP+FN}},
\end{eqnarray}
where TP, TN, FP, and FN denote the numbers of true positives, true negatives, false positives, and false negatives, respectively.
Based on the equations above, the ratio of precision to recall rates yields the correction factor to the intrinsic number of positive classes from that of model predictions ($\equiv\mathrm{(TP+FN)/(TP+FP)}$).
In addition, the area under the receiving operating curve (AU-ROC), and the area under the precision--recall curve (AU-PRC) are well-known evaluation metrics that are particularly useful in classification tasks based on imbalanced data (see section~\ref{s313}).
The AU-ROC is determined by the area below the receiver operating curve (ROC).
Here, the ROC is the plot of the true positive rate (TPR) versus the false positive rate (FPR) by varying the threshold, which is expressed as follows:
\begin{eqnarray} 
  \mathrm{TPR} &=& \frac{\mathrm{TP}}{\mathrm{TP+FN}} \\
  \mathrm{FPR} &=& \frac{\mathrm{FP}}{\mathrm{TN+FP}}.
\end{eqnarray}
The AU-ROC value is 0.5 in an ineffective model, whereas it approaches 1 in more complete models (figure~\ref{fig6}). 
Similarly, the AU-PRC is defined as the area under the precision--recall curve, and the value approaches one as the models become more ideal.
Table~\ref{tab2} indicates that the best trained model is sufficiently smart to differentiate between spiral and non-spiral galaxies because all metrics exceed 0.98 on the test data.

\begin{table}
\caption{Model evaluations in various metrics based on independent test data.
}
\begin{center}
\begin{tabular}{lrr}
    \hline
    Class     & Spiral galaxies & Ring galaxies\\ 
    \hline
    N\_positive & 1,000 & 100 \\ 
    N\_negative & 1,000 & 900 \\ 
    \hline
    Accuracy  & 0.984 & 0.958 \\
    Precision & 0.982 & 0.742 \\
    Recall    & 0.986 & 0.890 \\
    AU-ROC    & 0.999 & 0.987 \\
    AU-PRC    & 0.999 & 0.934 \\
    \hline
    \end{tabular}
\end{center}
\label{tab2}
\end{table}

\subsubsection{Ring classification}
\label{s313}

We trained the model classifier of the ring features as in the spiral classification.
However, the training process must be scrutinized because of the imbalanced data owing to the scarcity of ring galaxies. 
We indeed obtained only 930 ring galaxies (6.0\%) from the GC~DR1 catalog that satisfied the following criteria:
\begin{description}
\item[--] Ring: {\tt P(interaction)} $>0.5$ $\cap$ {\tt P(ring)} $>$ {\tt [P(fan), P(tail), P(distorted)]} $\to$ N$_\mathrm{positive}$=930
\item[--] Non-ring: {\tt P(spiral)} $<0.5$ not {\tt P(ring)} $>$ {\tt [P(fan), P(tail), P(distorted)]} $\to$ N$_\mathrm{negative}$=8,647.
\end{description}
These criteria have been determined through a discussion with the GALAXY CRUISE team and our visual inspection, to select ring galaxies in a rational manner.
Careless selection of ring galaxies for the training data lead to a useless ring classifier and hence different scientific results in section~\ref{s4} (e.g., no significant difference in stellar mass and SFR distributions between ring galaxies and the entire targets).
Examples of cutout images of randomly selected ring galaxies based on the criteria above are shown in figure~\ref{fig5}.
We selected 800 (100) ring galaxies and 7,200 (900) non-ring galaxies for the model training (and evaluation in table~\ref{tab2}).

Before conducting the model training, we employed several techniques that allowed the model to learn the imbalanced training data appropriately and, more specifically, to obtain better metric values for both the precision and recall rates, i.e., higher AU-PRC values.
Therefore, we modified the initial bias parameter ($b_0$) of the last sigmoid classifier (section~\ref{s311}) to improve its initial guess as follows:
\begin{equation}
b_0 \equiv \ln(N_\mathrm{positive}/N_\mathrm{negative}) = -2.197.
\end{equation}
Furthermore, the classifier was used to heavily weigh the ring samples by a factor of N$_\mathrm{negative}/$N$_\mathrm{positive}=9$ during the training epochs. 
The model with the highest validation AU-PRC value was selected as the best ring classifier through model training with early stopping.
These additional treatments allowed the model to focus more on under-represented ring classes, thereby realizing the construction of a better ring classifier.

\begin{figure}
\begin{center}
\includegraphics[width=8cm]{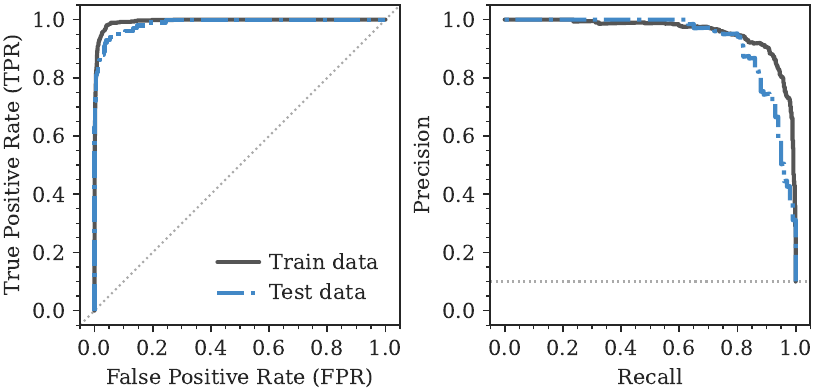}
\end{center}
\caption{
Left panel: ROC curve (TPR vs. FPR) of the trained classifier of the ring structure.
Solid black line and blue dash-dotted curve indicate ROC curves of training and test data, respectively.
Dotted diagonal gray line represents ROC curve if the model has no ability to classify the ring feature (i.e., AU-ROC $=0.5$).
Right panel: Precision--recall curve of the trained classifier of the ring structure.
Each line is the same as in the left panel.
}
\label{fig6}
\end{figure}

Consequently, we obtained the best model at the 19th epoch with the highest validation AU-PRC of 0.93 and validation accuracy of 0.97.
The AU-ROC and AU-PRC values obtained for the test data are shown in figure~\ref{fig6}.

\subsection{Morphological classifications}
\label{s32}

Based on the trained model dedicated to each dataset, we classified a total of 687,859 magnitude-limited galaxies at $z=$ 0.01--0.3 from HSC-SSP PDR3 (table~\ref{tab1}).
Figure~\ref{fig5} shows random examples of spiral and ring galaxies selected by the trained classifiers at different probability levels, which demonstrate their excellent capability in identifying spiral and ring structures.
Similar to previous studies \citep{Tadaki2020,Shimakawa2022}, the effectiveness of spiral classification was proven, as indicated by its high precision of 98\% on the test data (table~\ref{fig2}), although some misclassified objects were observed owing to, e.g., tidal features (figure~\ref{fig5}).
However, the model could not perform ring classification effectively as contamination was more critical due to the intrinsically small fraction of ring galaxies (6\% and 5\% in GC~DR1 and this study, respectively). 
Some ambiguous ring candidates were observed in randomly selected cutout images, particularly those with lower ring probabilities (figure~\ref{fig6}).
In fact, distinguishing ring features from spirals with low pitch angles, which are typically observed in passive spiral galaxies, is challenging and controversial, even for professional astronomers, as demonstrated by \citet{Shimakawa2022}.
Thus, catalog users must utilize the ring galaxy sample with caution, depending on the science case.

\begin{figure*}
\begin{center}
\includegraphics[width=16cm]{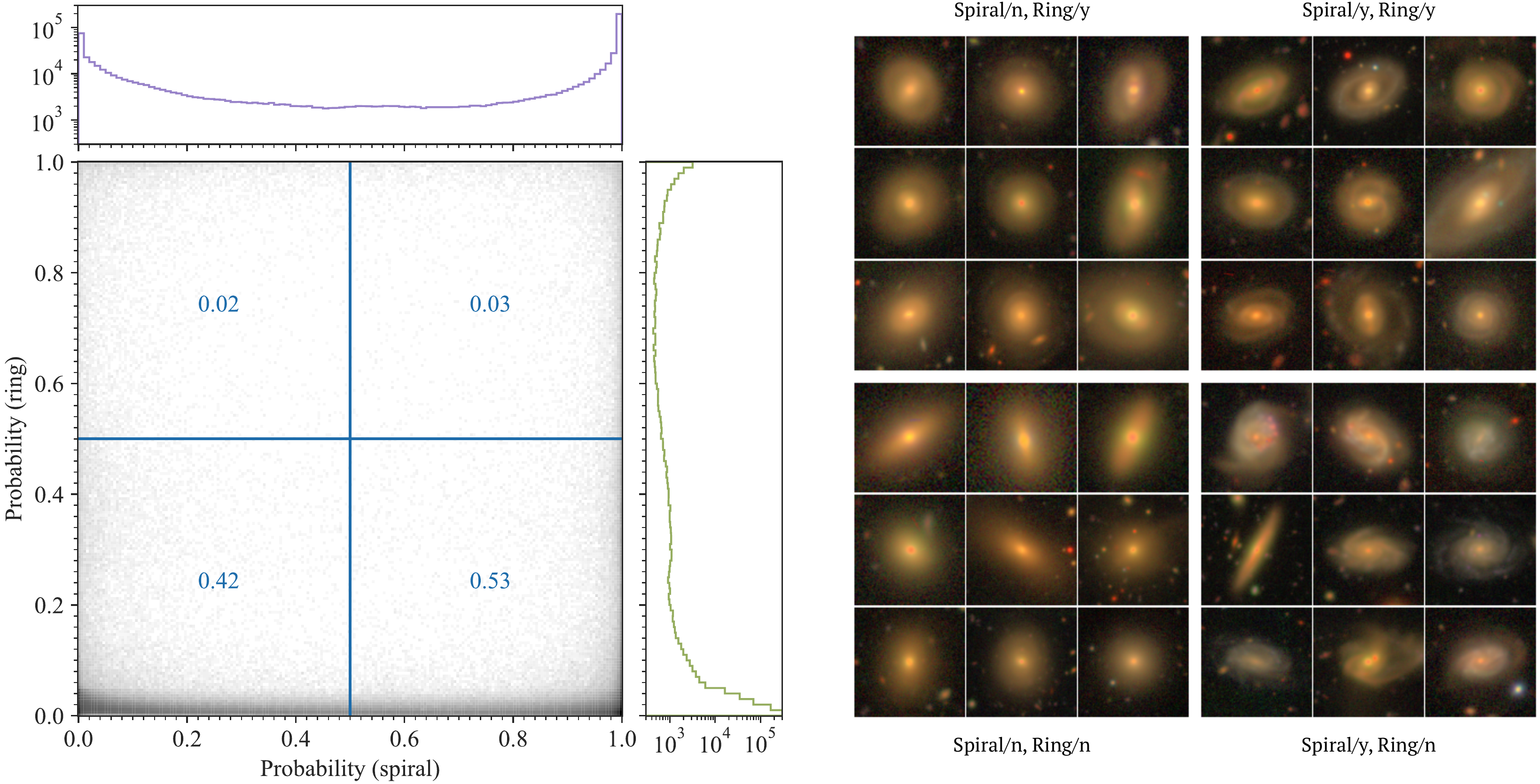}
\end{center}
\caption{
Left panel: Probability ($\hat{y}$) distribution in each class (table~\ref{tab1}).
Magenta and green histograms indicate those of spiral and ring probabilities, respectively.
Numbers in figure show fractions when separated by horizontal and vertical blue lines.
Right panel: Color postage stamps randomly selected from each group in the left panel.
The RGB color scaling follows that in {\tt hscMap} based on cutouts in the $i,r,g$-bands.
}
\label{fig7}
\end{figure*}

The resultant probability distribution ($\hat{y}$) in each classification task is presented in figure~\ref{fig7}, which shows that the models classified the targets accurately based on clear bimodal distributions in both the spiral and ring classifications.
This work separates the sample into two classes by $\hat{y}=0.5$ in each classification (the same threshold is applied thereafter).
Consequently, 385,449 spiral galaxies and 33,993 ring galaxies were selected, which constitute approximately 56\% and 5\% of the entire sample, respectively.
Table~\ref{tab1} summarizes the morphological classification results in the main sample and the subsample.
Among the ring galaxy sample, 54\% have both spirals and rings and the remaining 46\% are classified as ring galaxies without spirals.
They appeared differently in the four categories, as shown in figure~\ref{fig7}.
Broadly speaking, galaxies identified by the classifiers as having both spiral and ring features tend to be P-type rings, which feature crisp, knotty structures and often an offset nucleus \citep{Few1986}.
Otherwise, they can appear as outer pseudo-rings, which are more typically observed in late-type galaxies than early-type galaxies \citep{Elmegreen1992}.
By contrast, ring galaxies without spiral structures tend to exhibit smooth rings, which are labeled as O-type rings \citep{Few1986}.

\subsection{Morphology catalog}
\label{s33}

This section provides an overview of the source catalog and guidelines for using the morphology catalog, including some caveats.
The morphology catalog contains identification numbers, sky coordinates, and skymap ids in HSC-SSP PDR3; as well as object ids and redshifts in SDSS DR17 (table~\ref{tab3}).
Here, the skymap id is required to refer to the imaging qualities, such as the original seeing size and limiting magnitude at the target region in the HSC--SSP database \citep{Aihara2022}.
Our classification results, i.e., the prediction probabilities ($\hat{y}$) of the spiral and ring features, can be referenced by {\tt prob\_sp} and {\tt prob\_rg}, respectively.
Furthermore, the catalog provides the best-fit values and errors of stellar masses and SFRs for all targets to render it more useful.
They are derived from the SED fitting code, {\tt Mizuki} \citep{Tanaka2015}, as in the GC~DR1 paper \citep{Tanaka2023}.
The CModel magnitudes in the $ugriz$ bands from SDSS DR17 \citep{Abdurrouf2022} were adopted for the SED fitting by considering the flux loss issue of extended sources in the HSC-SSP pipeline (see section~\ref{s22}). 

\begin{table}
\tbl{Column description of the morphology catalog.}{%
\begin{tabular}{ll}
    \hline
    Column & Description\\ 
    \hline
    {\tt id} & object\_id in the HSC database\\
    {\tt id\_sdss} & object\_id in the SDSS DR17\\
    {\tt ra} & Right ascension ($^\circ$)\\
    {\tt dec} & Declination ($^\circ$)\\
    {\tt skymap\_id} & skymap\_id in the HSC database\\
    {\tt sample} & sample class (main or sub; table~\ref{tab1})\\
    {\tt z} & redshift\\
    {\tt z\_flag} & redshift flag (spec or phot)\\
    {\tt prob\_sp} & spiral probability\\
    {\tt prob\_rg} & ring probability\\
    {\tt reduced\_chisq} & reduced $\chi^2$ in the Mizuki SED fitting\\
    {\tt mstar} & stellar mass (M$_\odot$) from Mizuki\\
    {\tt mstar\_err68min} & lower 68\% confidence limit of {\tt mstar}\\
    {\tt mstar\_err68max} & upper 68\% confidence limit of {\tt mstar}\\
    {\tt sfr} & SFR (M$_\odot$yr$^{-1}$) from Mizuki\\
    {\tt sfr\_err68min} & lower 68\% confidence limit of {\tt sfr}\\
    {\tt sfr\_err68max} & upper 68\% confidence limit of {\tt sfr}\\
    \hline
    \end{tabular}}
\label{tab3}
\end{table}

\begin{figure}
\begin{center}
\includegraphics[width=8cm]{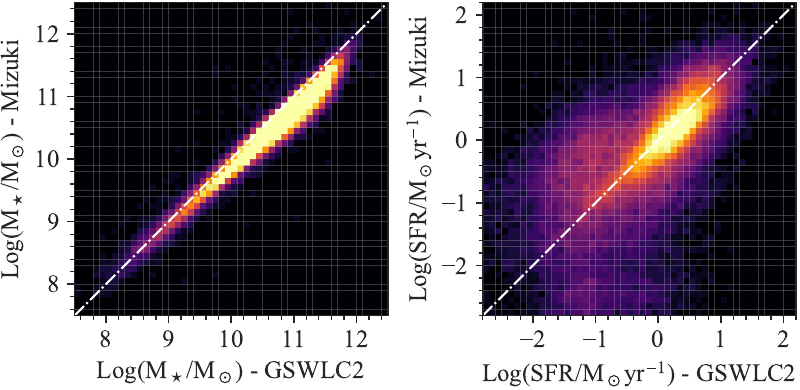}
\end{center}
\caption{
Comparisons of derivations of stellar masses (left) and SFRs (right) between Mizuki \citep{Tanaka2015} and GSWLC-X2 \citep{Salim2018} for 63,700 GSWLC-X2 sources included by HSC-SSP PDR3.
Dash-dotted white lines show 1:1 relation.
}
\label{fig8}
\end{figure}

\begin{figure}
\begin{center}
\includegraphics[width=7.5cm]{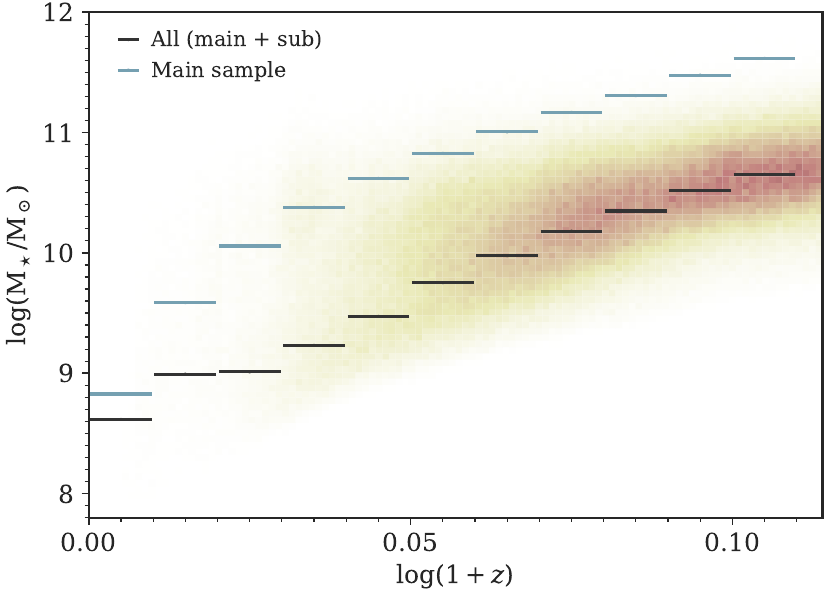}
\end{center}
\caption{
Logarithmic stellar mass vs. redshift in our sample.
Black and blue lines show the 90\% stellar mass completeness limits at each redshift bin in the entire sample and main sample, respectively.
}
\label{fig9}
\end{figure}

For validation, we compared the stellar masses and SFRs from {\tt Mizuki} in GC~DR1 with those from the GALEX--SDSS--WISE Legacy Catalog (GSWLC-X2; \cite{Salim2016,Salim2018}), as shown in figure~\ref{fig8}.
In the GSWLC-X2 catalog, the {\tt CIGALE} SED fitting code \citep{Burgarella2005,Noll2009,Boquien2019} was used for bright galaxies with a Petrosian $r$-band magnitude of $<18$ mag at $z_\mathrm{spec}=$ 0.01--0.3, based on the multi-wavelength datasets of the GALEX GR6/7 \citep{Martin2005}, SDSS DR7 \citep{Abazajian2009}, and WISE photometry \citep{Lang2016} 
Therefore, we assume that their obtained stellar masses and SFRs are more reliable than our measurements, which rely solely on the SDSS $ugriz$-band photometry for managing fainter objects down to $r=20$ mag.
Figure~\ref{fig8} shows that the two stellar mass measurements were consistent, although a systematic offset was indicated at $\sim0.2$ dex, which can occur when different model assumptions are adopted, such as the stellar populations, star formation histories, and datasets.
For instance, GSWLC-X2 optimized the age of the main stellar population to 10~Gyr for local galaxies \citep{Salim2016}, whereas such optimization was not indicated in the {\tt Mizuki} measurements, which might have resulted in the systematic difference.
The 90\% stellar mass completeness limit was calculated based on the prescription by \citet{Pozzetti2010}.
The obtained completeness limits at binned redshift ranges on a logarithmic scale are shown in figure~\ref{fig9}. 
The findings suggest that the main sample (and the combined dataset with the subsample) is mass complete down to M$_\star\sim10^{10.6}$ ($10^{9.4}$), $10^{11.2}$ ($10^{10.2}$), and $10^{11.6}$ ($10^{10.7}$) M$_\odot$ at $z=0.1$, 0.2, and 0.3, respectively.
However, the obtained SFRs were largely scattered, particularly at lower SFRs compared with those of the GSWLC-X2 catalog.
Such a large scatter is expected because we did not adopted GALEX and WISE photometry, which are important for measuring accurate SFRs.

\begin{figure}
\begin{center}
\includegraphics[width=7.5cm]{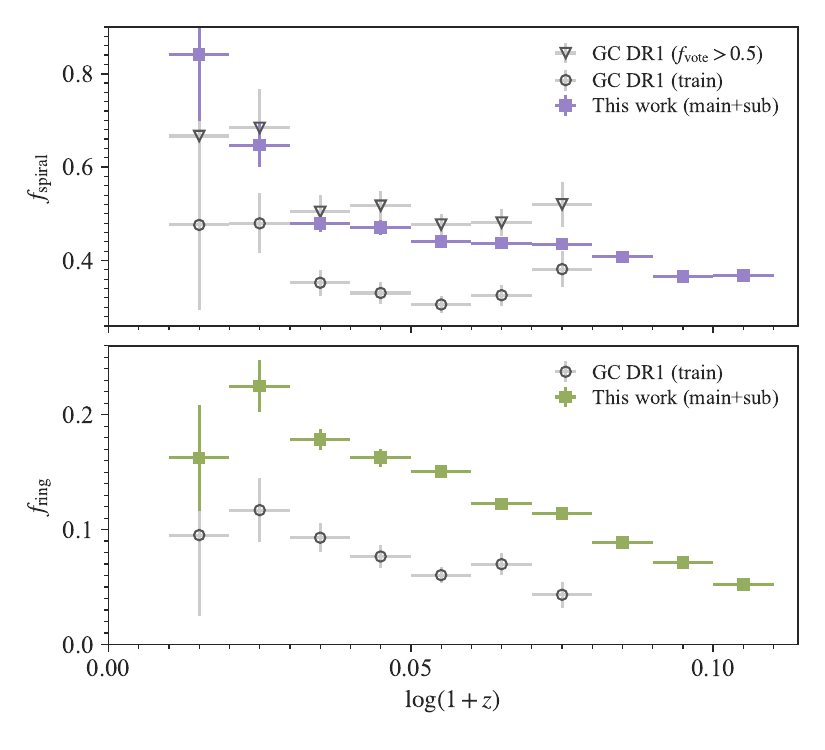}
\end{center}
\caption{
Top panel: Spiral fractions for the entire sample with M$_\star=10^{10.7}$--$10^{11}$ M$_\odot$ at different redshift bins.
Magenta squares show spiral fractions in our sample. 
Inverse black triangles and circles are those from GC~DR1 samples with the selection criteria {\tt galaxy\_type} $>0.5$ and $>0.79$ (same as for the training data), respectively.
Error-bars indicate $1\sigma$ Poisson noises.
Bottom panel: Same as the top panel, but for ring fractions.
Our sample is shown in green color.
}
\label{fig10}
\end{figure}

We verified whether our morphological classifications had any redshift bias by comparing the spiral and ring fractions ($f_\mathrm{spiral}$ and $f_\mathrm{ring}$, respectively), as a function of redshift (figure~\ref{fig10}). 
In this regard, we adopted galaxies with at narrow stellar mass range of M$_\star=10^{10.7}-10^{11}$ M$_\odot$ for a fair comparison.
We confirmed that the spiral fraction based on our deep learning classification indicated a values between those from GC~DR1 with two different selection criteria, i.e., $f_\mathrm{vote}$ ({\tt galaxy\_type} in the GC~DR1 catalog) $>0.5$ and $>0.79$, where the latter threshold was used for selecting the training data according to suggestions from \citet{Tanaka2023}.
Similarly, the obtained ring fraction is higher than that based on the training sample because they are selected by prioritizing accuracy over completeness (section~\ref{s313}).
Notably, redshift dependencies can be seen in both our targets and the GC~DR1 samples, thus suggesting that they are not caused solely by a model-sensitive issue.
At very low redshift ($z<0.07$), the main factor of the apparent redshift trends of the morphological fractions would be the cosmic variance, owing to the small cosmological volume coverage ($2.4\times10^{-3}$ Gpc$^3$) at $z=$ 0.01--0.07 (see also discussion in \cite{Tanaka2023}).
Besides, some diffuse tidal features and inner halos of local bright objects are misidentified as spiral structures based on further visual verification, as discussed in the previous subsection.
However, the decreasing trend of the spiral fraction at $z>0.2$ would be due to the sampling bias of the training data from GC~DR1, which is limited to the spec-$z$ sources at $z=$ 0.01--0.2.
Such a sampling bias affected the ring classification as well, as a similar decreasing trend can be observed in the ring fraction (figure~\ref{fig8}).
In the case of the ring classification, the physical spatial resolution of $\sim1$ kpc or less is preferable to separate the inner rings from central bulges and spiral arms, which can be achieved at $z\lesssim0.1$ under the typical seeing condition of $\sim0.6$ arcsec.
This resolution limit further results in redshift dependence, as observed in both our targets and the GC~DR1 sample.

\subsubsection*{Tips on how to use the catalog}

The catalog provides the classification results in the entire sample (table~\ref{tab1}) to allow for some flexibility.
However, in light of these issues above, we provide some recommendations for the selection thresholds in the morphology catalog (table~\ref{tab3}) to obtain relatively reliable spiral and ring galaxies for different purposes.
First, limiting to the main sample at $z=$ 0.07--0.2 or a narrower redshift slice is encouraged to avoid cosmic variance and classification bias (figure~\ref{fig10}).
Ring classification is more sensitive to the redshift bias because the spatial resolution is particularly critical for resolving inner ring features.
Users may adopt the subsample at their own risk. 
The subsample may be useful to increase the sample size depending on the science case, they would undersample spiral and ring galaxies due to the classification bias.
Additionally, one can establish more reliable spiral and ring galaxy samples by applying a more strict probability cut (e.g., $\hat{y}>0.95$). 
Our SFR measurements indicated significant uncertainties (figure~\ref{fig8}); therefore, they must be used with caution.
For a mass-complete sample within a certain redshift slice, readers can refer to figure~\ref{fig9}.

\section{Discussion}
\label{s4}

In the following sections, we discuss the morphological dependencies of the stellar mass, SFR, and cluster environments.
The primary aim of this study is to investigate ring galaxies because they are investigated less than spiral galaxies.
We hereafter use only the main sample with $r<17.8$ mag (table~\ref{tab1}), unless otherwise noted, to minimize a potential classification bias due to the limited training dataset from GC~DR1 (section~\ref{s21}).

\subsection{Stellar mass dependence}
\label{s41}

\begin{figure}
\begin{center}
\includegraphics[width=8cm]{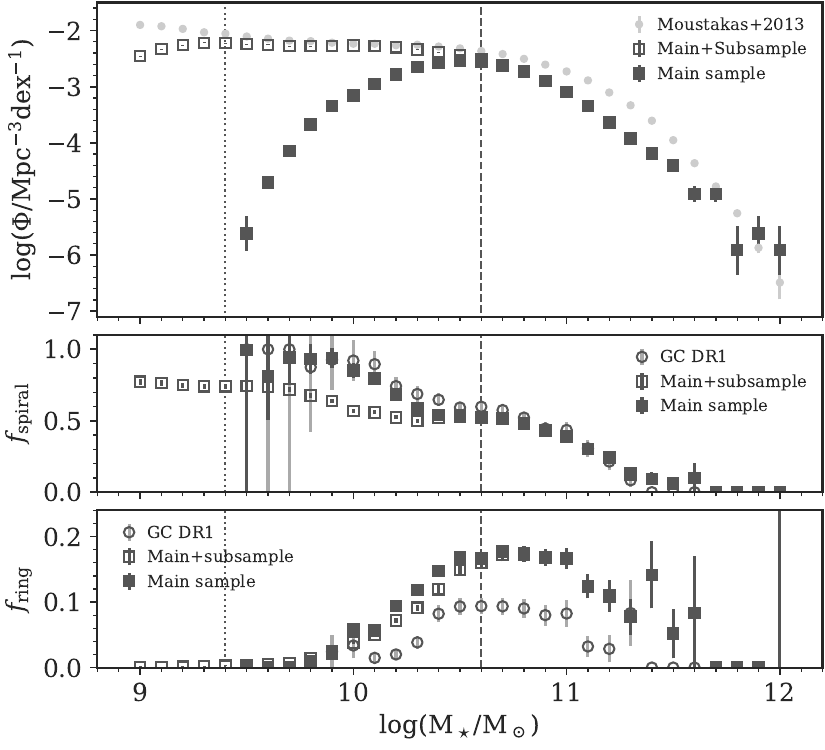}
\end{center}
\caption{
Top panel: Logarithmic stellar mass function at $z=0.10\pm0.02$.
Filled and open squares show the stellar mass functions for the main sample and the entire sample, respectively.
Figure also shows stellar mass function at $z=$ 0.01--0.2 by \citet{Moustakas2013} with gray circles for reference.
The vertical dashed and dotted lines indicate the mass completeness limits of the main and the entire sample, respectively (see also figure~\ref{fig9}).
Middle and bottom panels show fractions of spiral and ring galaxies as a function of logarithmic stellar mass at $z=0.10\pm0.02$.
Black circles indicate morphological fractions in GC~DR1, where spiral galaxies are selected via {\tt galaxy\_type} $>0.5$.
Error-bars in each panel depict $1\sigma$ Poisson errors.
}
\label{fig11}
\end{figure}

First, we discuss the spiral and ring fractions for a given stellar mass based on the obtained stellar mass information.
Figure~\ref{fig11} shows the stellar mass function as well as the morphological fractions of spiral and ring galaxies based on the HSC-SSP PDR3 sample at $z=0.10\pm0.02$ ($N=$ 18,017).
The redshift range was determined to reduce the redshift bias and sufficiently encompass the stellar mass down to $\sim10^{10}$ M$_\odot$.
As reported in previous studies (e.g., \cite{Bamford2009,Tanaka2023}), the spiral fraction decreased with increasing stellar mass. 
It should be noted that the main and GC~DR1 sample are biased towards star-forming galaxies with low mass--to-light ratios for M$_\star\lesssim10^{10.6}$ M$_\odot$ at $z=0.10\pm0.02$ (figure~\ref{fig9}).
Indeed, their spiral fractions are significantly higher than that from our combined sample of the main and subsample at M$_\star\lesssim10^{10.6}$ M$_\odot$ below the mass completeness limit, while the combined sample may somehow undersample spiral galaxies due to the possible classification bias.

By contrast, the ring fractions show an intriguing trend against the stellar mass (figure~\ref{fig11}), i.e., peak fractions of $\sim0.18$ at M$_\star\sim10^{10.5}$--$10^{11}$ M$_\odot$ and almost zero fractions are indicated at both the low-mass and high-mass ends.
Such a stellar mass dependence seems not to be caused by the selection bias because we see the same trend in the combined data with the subsample.
A similar trend can be observed in the training data from the GC~DR1 catalog (figure~\ref{fig11}) and cosmological hydrodynamic simulations \citep{Snyder2015,Elagali2018}, thus further supporting this peculiar trend.
The peak fractions at M$_\star\sim10^{10.5}$--$10^{11}$ M$_\odot$ would be related to the increasing fraction of barred disc galaxies \citep{Masters2012} and green valley galaxies (see section~\ref{s42}) at this stellar mass range.
The fraction excess may be due to galaxies above M$_\star=10^{10.5}$, which generally reside in massive halos $>10^{12}$ M$_\odot$ (e.g., \cite{vandenBosch2007,Zu2015,Correa2020}), thus indicating that they would be surrounded by galaxies that can trigger ring structures (e.g., \cite{Appleton1996,Romano2008,Elagali2018}).
However, it should be noted that the resolution limit may affect the turnover mass, because we confirmed that this value gets slightly higher (lower) when we used the samples at higher (lower) redshift bins.
A high-resolution wide-field survey with space telescopes, such as Euclid \citep{Laureijs2011,Amendola2013} and Roman Space Telescope \citep{Spergel2015} may be required to fully address this possible tendency.

\subsection{Star formation dependence}
\label{s42}

\begin{figure*}
\begin{center}
\includegraphics[width=16cm]{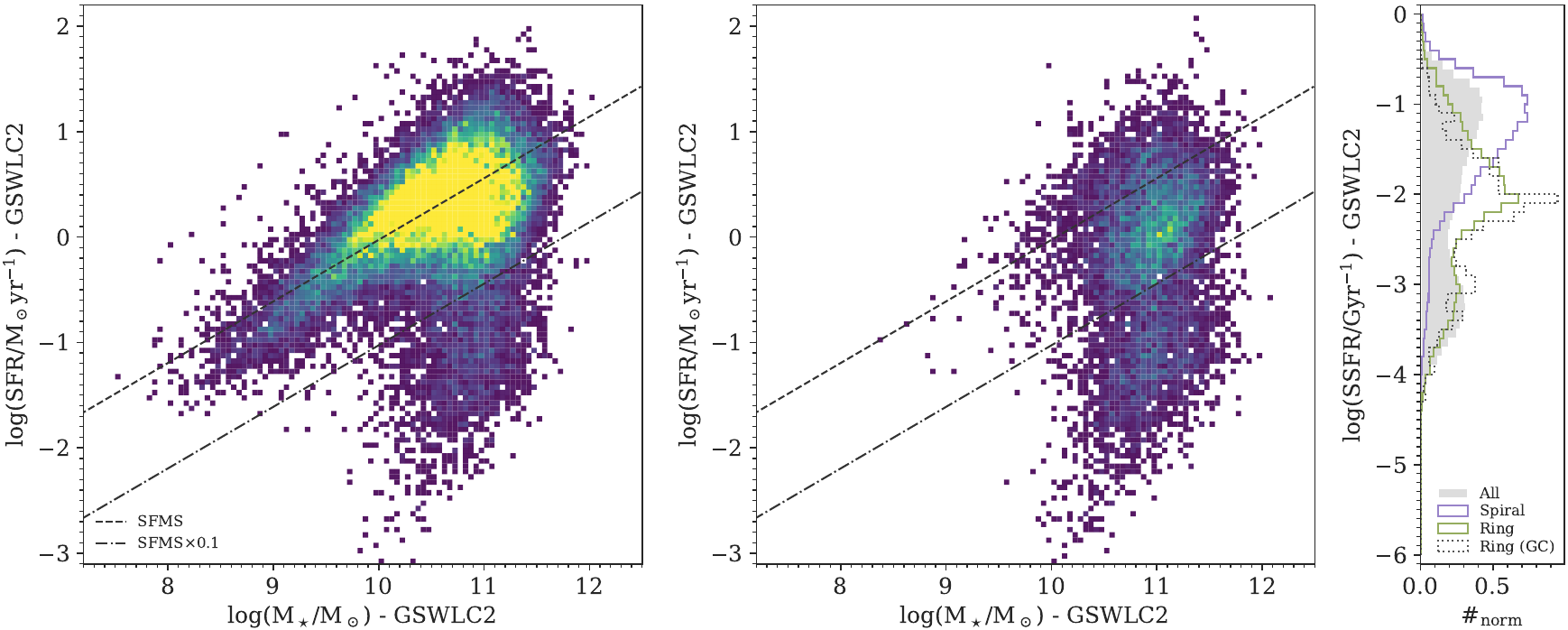}
\end{center}
\caption{
From left to right, SFR vs. stellar mass on logarithmic scale for spiral and ring galaxies, and normalized SSFR distributions based on GSWLC-X2 catalog \citep{Salim2018}.
Dashed black lines are best-fit star-forming main sequence for each sample, and values 1 dex below these lines are depicted by dash-dotted black lines. 
Gray-filled, magenta, green, and dotted black histograms on right panels show the main sample, spirals, ring galaxies from this work, and ring galaxies from GC~DR1 via the selection threshold in section~\ref{s313}, respectively.
}
\label{fig12}
\end{figure*}

In this subsection, we discuss the morphological dependence of the star-forming main sequence \citep{Brinchmann2004,Salim2007}, particularly regarding the obtained ring galaxies.
We tested star formation properties using independently measured stellar masses and SFRs from the {\tt CIGALE} SED fitting code by \citet{Salim2018} (see also section~\ref{s33}).
Notably, the \citet{Salim2018} estimations are limited to 59,854 spec-$z$ sources with $r<17.8$ mag in our main sample, among which the spiral and ring galaxies are limited to 31,864 and 8,808 sources, respectively.

The resultant SFR vs. stellar mass planes are shown in figure~\ref{fig12}.
The selected spiral galaxies were preferentially located on the star-forming main sequence, as commonly reported in previous studies (e.g., \cite{Willett2015,Wang2020,Shimakawa2022}).
By contrast, we confirmed that ring galaxies preferentially resided in the green valley between the star-forming main sequence and passive sequence.
The same trend was also confirmed in GC~DR1 but for the limited sample ($N=906$).
Specifically, when we limited the main sample to only massive galaxies with M$_\star=10^{10.5}$--$10^{11}$ M$_\odot$ at $z=0.10\pm0.02$, the ring fraction increased from $19.0\pm0.5\%$ to $33.7\pm1.9\%$ at specific SFR (SSFR) $=10^{-2.0\pm0.2}$ Gyr$^{-1}$.

The obtained trend is consistent with a previous study pertaining to mass-selected galaxies with M$_\star=10^{10.25}$--$10^{10.75}$ M$_\odot$ at $z<0.075$ \citep{Smith2022}, which demonstrated an excess of the ring fraction in the green valley on a color--stellar mass diagram based on the Galaxy Zoo classification.
In addition, our results broadly agree with theoretical predictions from a cosmological hydrodynamic simulation \citep{Elagali2018}.
\citet{Elagali2018} reported that the star formation efficiency of ring galaxies tend to be low primarily owing to the low ISM pressure and metallicity in the outer regions, which may be caused by the acquisition of metal-poor gas from the dwarf via minor mergers.
Although we tentatively verified whether they exhibited lower gas-phase metallicities using the MPE-JHU catalog \citep{Tremonti2004}, we did not identify such metallicity dependence.
This might be because the metallicity measurement performed was based on spectra obtained from SDSS DR7 spectroscopy involving fibers of 3 arcsec diameter \citep{Abazajian2009}, which implies that most of the line fluxes should originate from the central regions.
A systematic and comprehensive survey with integral field units is required to investigate the metallicity dependence of galaxy outskirts.

\begin{figure}
\begin{center}
\includegraphics[width=8cm]{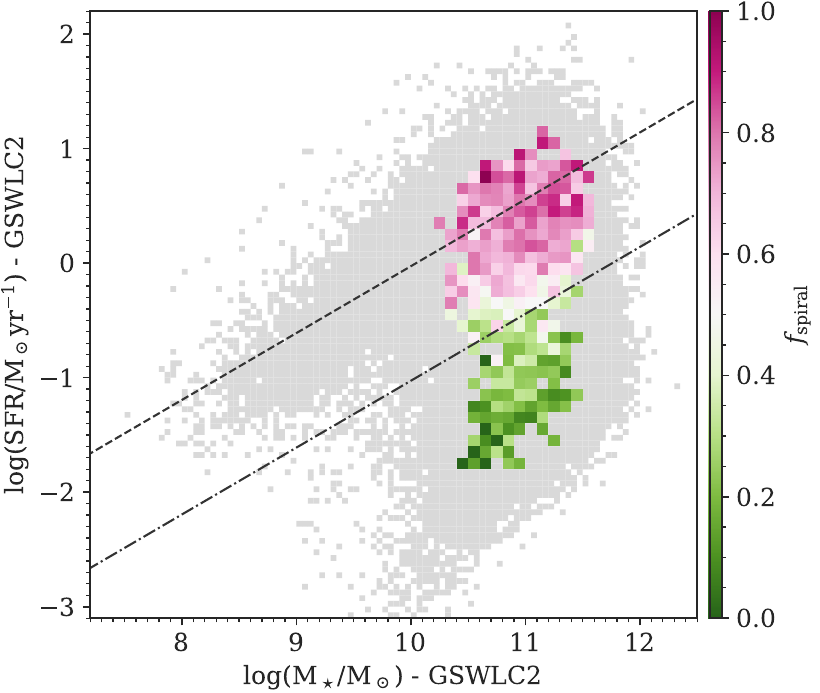}
\end{center}
\caption{
Same as top middle panel of figure~\ref{fig12}, except that colored region indicates spiral fraction (with {\tt prob\_sp > 0.5} in table~\ref{tab3}) in each SFR--M$_\star$ bin, as shown by color bar on the right.
Gray area shows all GSWLC-X2 counterparts.
}
\label{fig13}
\end{figure}

We obtained 18,441 galaxies (or 4,789 in the main sample) with both spiral and ring features, which constituted 54\% of the entire ring galaxy sample (or 58\% in the main sample).
Additionally, we investigated their star formation phases with and without spiral features (figure~\ref{fig13}).
The spiral fractions obtained from ring galaxies clearly depend on star formation, as the star-forming (or passive) ring galaxies are dominated by spiral (or non-spiral) galaxies.
This trend is also consistent with our visual impression (figure~\ref{fig7}) that ring galaxies with spiral arms seem more P-types, while those without spirals tend to show O-type ring features (section~\ref{s32}).
Such star formation dependence may suggest a morphological transition linked to the evolutionary stage on the galactic disk.

\subsection{Environmental dependence}
\label{s43}

We revisited the morphology--density relation in cluster environments based on the main sample at $z=$ 0.07--0.2.
It is commonly known that the spiral fraction decreases at higher densities \citep{vanderWel2008,Tasca2009,Calvi2012} and towards the center of galaxy clusters (\cite{Dressler1980,Larson1980,Postman2005,Vulcani2011}, and see also more comparative considerations in \cite{Bamford2009,Vulcani2023}).
Therefore, we were particularly interested in investigating whether the ring fraction changes in cluster environments, which has not been investigated sufficiently hitherto (e.g., \cite{Elmegreen1992,Fernandez2021}).

\begin{figure}
\begin{center}
\includegraphics[width=7.5cm]{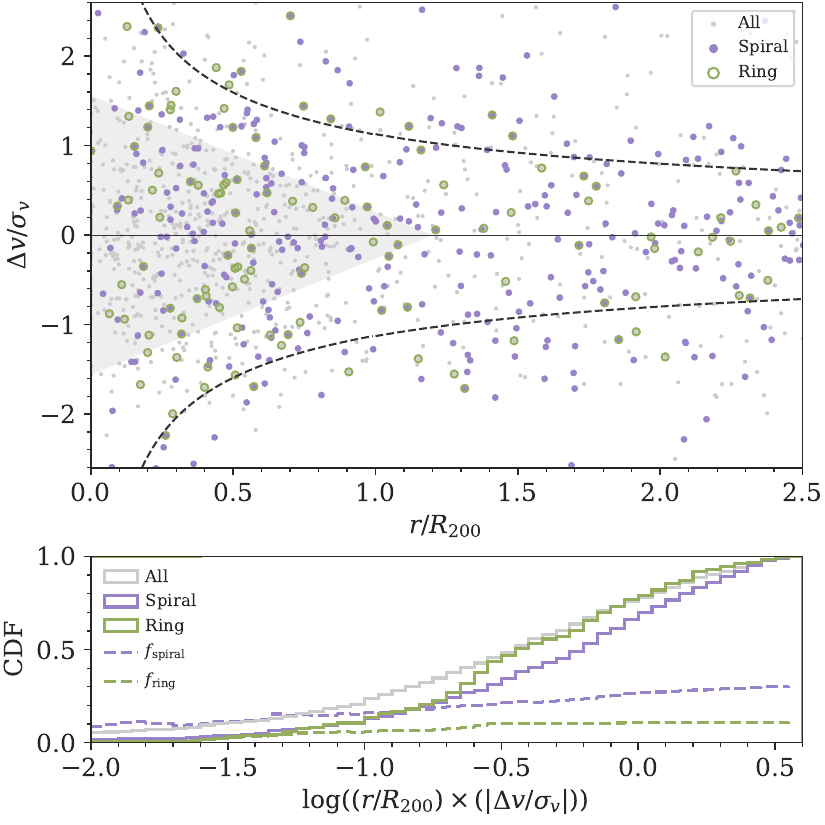}
\end{center}
\caption{
Top panel: Galaxy distributions on phase--space diagram.
Circles indicate member galaxies of X-ray clusters \citep{Clerc2016} with $z_\mathrm{spec}=$ 0.07--0.2, where spiral and ring galaxies are shown in magenta and green, respectively.
Dashed curves indicate escape velocity, and gray region shows virialized area based on \citet{Jaffe2015}.
Bottom panel: Cumulative distribution functions (CDFs) on accretion phase ($(R/R_{200})\times(|\Delta v/\sigma_v|)$) in each morphological class.
Gray, magenta, and green lines indicate the entire sample, spiral galaxy, and ring galaxy, respectively.
Colored dashed histograms show spiral fraction (magenta) and ring fraction (green) given the accretion phase.
}
\label{fig14}
\end{figure}

Herein, the morphological dependence on the cluster environments is discussed based on phase--space and accretion phase diagrams with a scaled projected radius ($r/R_{200}$) and peculiar velocity ($\Delta v/\sigma$), where $R_{200}$ and $\sigma$ are the cluster virial $R_{200}$ radius and velocity dispersion, respectively.
These indices allow us to investigate galaxy populations in a given accretion state without being biased by the projected position from the cluster center in a (semi-)theoretical manner \citep{Noble2013,Noble2016,Jaffe2015,Muzzin2014,Hayashi2017}.
We adopted the SPectroscopic IDentification of eROSITA Sources (SPIDERS; \cite{Clerc2016}), which constitute the SDSS-IV project \citep{Blanton2017}, to obtain X-ray selected clusters with $\sim10^{14}$--$10^{15}$ M$_\odot$ associated with our targets.
In particular, we selected 41,978 massive galaxies with M$_\star>1\times10^{10}$ M$_\odot$ at $z_\mathrm{spec}=$ 0.07--0.2 to reduce the selection bias (section~\ref{s33}), and then cross-matched them with the SPIDERS cluster catalog.
Consequently, we obtained 733 sources, including 256 spiral and 110 ring galaxies, which were successfully matched to 70 X-ray clusters within the logarithmic accretion phase of $\log((r/R_{200})\times(|\Delta v/\sigma_v|))<0.6$ (figure~\ref{fig14}). 
The spiral and ring fractions were estimated to be $34.9\pm2.5\%$ and $15\pm1.5\%$ at M$_\star>1\times10^{10}$ M$_\odot$, which were lower than the mean fractions of $49.0\pm0.4\%$ (20,580/41,978) and $17.2\pm0.2\%$ (7,211/41,978), respectively.

Figure~\ref{fig14} shows the manner by which the spiral and ring structures of the cluster members diminished during their infalling phases.
As shown by the cumulative distribution functions (CDFs) of the entire sample, the spiral and ring galaxies indicated decreasing spiral and ring fractions towards the cluster center, which were statistically significant based on the Kolmogorov--Smirnov test ($p<0.05$) at $\log((r/R_{200})\times(|\Delta v/\sigma_v|))\lesssim0$.
The decreasing spiral fraction in the local galaxy cluster has been investigated extensively as a morphology--density relation (e.g., \cite{Dressler1980,Larson1980,Vulcani2023}); thus, this specific details are omitted herein.
Although some spiral galaxies remain within the virialized region, approximately 40\% are passive, as reported by \citet{Shimakawa2022}.
Similarly, the ring fraction declines towards the cluster center, particularly at $\log((r/R_{200})\times(|\Delta v/\sigma_v|))<-0.5$ (figure~\ref{fig14}), supporting the modern hydrodynamic cosmological simulation \citep{Elagali2018}.
The formation mechanisms of the ring features can be classified into two scenarios, i.e., galactic interactions and gravitational effects induced by a central galactic bar (e.g., \cite{Lynds1976,Appleton1996,Elagali2018}).
As both cases require galactic disks, the decreasing ring fraction in the cluster center can be naturally explained by the lack of disk galaxies in the galaxy clusters, which is known as the morphology--density relation.
Similarly, \citet{Elmegreen1992} reported decreasing fractions of late-type galaxies with outer rings in local galaxy clusters.

Finally, we should note that the ring galaxy sample is likely to be contaminated by pseudo-ring spiral galaxies. 
Based on the currently available imaging data, the ring structure cannot be fully distinguished from spiral arms with a small pitch angle (see also \cite{Shimakawa2022}), which would cause uncertainties in measuring the intrinsic number densities and fractions of ring galaxies beyond Poisson errors.
We leave this caveat to future work with high-resolution imaging data from the wide-field surveys in space (e.g., \cite{Laureijs2011,Amendola2013,Spergel2015}).

\section{Conclusion}
\label{s5}

In this study, spiral and ring classifications were performed for 687,859 galaxies down to the $r$-band magnitude of 20 mag, which comprised 59,854 main sample and 628,005 subsample at $z=$ 0.01--0.3 over 930 deg$^2$ in the effective area.
We adopted photometry and redshift information from SDSS DR17 \citep{Blanton2017,Abdurrouf2022} and imaging data from the HSC-SSP PDR3 Wide layer \citep{Aihara2022}. 
To effectively classify the 687,859 images, we employed visual morphology classifications from the GALAXY CRUISE citizen science project, which is dedicated to HSC imaging data \citep{Tanaka2023}.
We then constructed labeled training, validation, and test datasets to train deep learning models for spiral and ring classifications.

Consequently, we obtained 31,864 spiral galaxies and 8,808 ring galaxies for the main sample (385,449 and 33,993 when combined with the subsample, respectively), which constituted 53\% and 15\% of the targets, respectively.
The trained deep learning classifiers indicated AU-PRC values of 0.999 and 0.934 for the test data (table~\ref{tab2}).
In addition to the prediction probabilities of the spiral and ring features, our morphology catalog provided stellar masses and SFRs for all objects from the {\tt Mizuki} SED fitting code, as in \citet{Tanaka2023}.
The full source catalog is available on the GALAXY CRUISE website.

The large morphology catalog enables us to statistically examine the general properties of spiral and ring galaxies.
Based on the main sample combined with the GSWLC-X2 catalog \citep{Salim2018}, we confirmed the increasing spiral fraction at lower mass galaxies, and more interestingly, we observed a plateau in the ring fraction at a particular stellar mass range of M$_\star=10^{10.5}$--$10^{11}$ M$_\odot$.
Also, our results show that spiral galaxies mostly follow the star-forming main sequence, whereas ring galaxies are preferentially located in the green valley at M$_\star=10^{10.5}$--$10^{11}$ M$_\odot$.
Furthermore, we observed fading spiral features of ring galaxies from the star-forming main sequence to passive galaxies, which implied their morphological transitions across the green valley.
In cluster environments, the fractions of both spiral and ring galaxies decrease toward the cluster center owing to the decreasing fraction of disk galaxies, which is known as the morphology--density relation.

However, it should be noted that our classifications, particularly for ring galaxies, presented sampling and selection biases due to various limitations such as limited survey volume and imaging qualities. 
In addition, the ring classification must be further improved to distinguish between inner and outer rings, and between pseudo-rings, such that we can determine the formation mechanisms of various types of ring structures in more detail.
This study provides recommendations for alleviating issues depending on science use, as summarized in section~\ref{s33}.
Further research is required to comprehensively analyze the (recurrent) formation and dissipation mechanisms of spiral and ring structures, such as follow-up integral field spectroscopy and submillimeter observations, to resolve the kinematics and cold gas properties. 
The ongoing Euclid Wide Survey with the VIS instrument \citep{EuclidCollaboration2022} will help us determine ring structures more precisely.


\begin{ack}
The morphological classification catalog is publicly available on the GALAXY CRUISE website (\url{https://galaxycruise.mtk.nao.ac.jp/en/}).
Based on data collected at the Subaru Telescope and retrieved from the HSC data archive system, which is operated by Subaru Telescope and Astronomy Data Center at National Astronomical Observatory of Japan. We are honored and grateful for the opportunity of observing the Universe from Maunakea, which has the cultural, historical and natural significance in Hawaii. 

The Hyper Suprime-Cam (HSC) collaboration includes the astronomical communities of Japan and Taiwan, and Princeton University. The HSC instrumentation and software were developed by the National Astronomical Observatory of Japan (NAOJ), the Kavli Institute for the Physics and Mathematics of the Universe (Kavli IPMU), the University of Tokyo, the High Energy Accelerator Research Organization (KEK), the Academia Sinica Institute for Astronomy and Astrophysics in Taiwan (ASIAA), and Princeton University. Funding was contributed by the FIRST program from Japanese Cabinet Office, the Ministry of Education, Culture, Sports, Science and Technology (MEXT), the Japan Society for the Promotion of Science (JSPS), Japan Science and Technology Agency (JST), the Toray Science Foundation, NAOJ, Kavli IPMU, KEK, ASIAA, and Princeton University. 
This paper makes use of software developed for the Large Synoptic Survey Telescope. We thank the LSST Project for making their code available as free software at \url{http://dm.lsst.org}.

The Pan-STARRS1 Surveys (PS1) have been made possible through contributions of the Institute for Astronomy, the University of Hawaii, the Pan-STARRS Project Office, the Max-Planck Society and its participating institutes, the Max Planck Institute for Astronomy, Heidelberg and the Max Planck Institute for Extraterrestrial Physics, Garching, The Johns Hopkins University, Durham University, the University of Edinburgh, Queen’s University Belfast, the Harvard-Smithsonian Center for Astrophysics, the Las Cumbres Observatory Global Telescope Network Incorporated, the National Central University of Taiwan, the Space Telescope Science Institute, the National Aeronautics and Space Administration under Grant No. NNX08AR22G issued through the Planetary Science Division of the NASA Science Mission Directorate, the National Science Foundation under Grant No. AST-1238877, the University of Maryland, and Eotvos Lorand University (ELTE) and the Los Alamos National Laboratory.

This paper is based in part on data from the Hyper Suprime-Cam Legacy Archive (HSC~LA), which is operated by the Subaru Telescope. The original data in HSC~LA was collected at the Subaru Telescope and retrieved from the HSC data archive system, which is operated by the Subaru Telescope and Astronomy Data Center at National Astronomical Observatory of Japan. 

Funding for the Sloan Digital Sky Survey IV has been provided by the Alfred P. Sloan Foundation, the U.S. Department of Energy Office of Science, and the Participating Institutions. 
SDSS-IV acknowledges support and resources from the Center for High Performance Computing at the University of Utah. The SDSS website is \url{www.sdss.org}.

SDSS-IV is managed by the Astrophysical Research Consortium for the Participating Institutions of the SDSS Collaboration including the Brazilian Participation Group, the Carnegie Institution for Science, Carnegie Mellon University, Center for Astrophysics | Harvard \& Smithsonian, the Chilean Participation Group, the French Participation Group, Instituto de Astrof\'isica de Canarias, The Johns Hopkins University, Kavli Institute for the Physics and Mathematics of the Universe (IPMU) / University of Tokyo, the Korean Participation Group, Lawrence Berkeley National Laboratory, Leibniz Institut f\"ur Astrophysik Potsdam (AIP), Max-Planck-Institut f\"ur Astronomie (MPIA Heidelberg), Max-Planck-Institut f\"ur Astrophysik (MPA Garching), Max-Planck-Institut f\"ur Extraterrestrische Physik (MPE), National Astronomical Observatories of China, New Mexico State University, New York University, University of Notre Dame, Observat\'ario Nacional / MCTI, The Ohio State University, Pennsylvania State University, Shanghai Astronomical Observatory, United Kingdom Participation Group, Universidad Nacional Aut\'onoma de M\'exico, University of Arizona, University of Colorado Boulder, University of Oxford, University of Portsmouth, University of Utah, University of Virginia, University of Washington, University of Wisconsin, Vanderbilt University, and Yale University.

We would like to thank Editage (\url{www.editage.com}) for English language editing.
This research was financially supported by JSPS KAKENHI Grant Number 22H01270 and 22K14078.
This work made extensive use of the following tools, {\tt NumPy} 
\citep{Harris2020}, {\tt Matplotlib} \citep{Hunter2007}, the Tool for OPerations 
on Catalogues And Tables, {\tt TOPCAT} \citep{Taylor2005}, a community-developed 
core Python package for Astronomy, {\tt Astopy} \citep{AstropyCollaboration2013}, 
and Python Data Analysis Library {\tt pandas} \citep{Team2022}. 
\end{ack}


\bibliographystyle{pasj}
\bibliography{rs23c} 

\end{document}